\documentclass[referee]{raa_rb}

\usepackage{graphicx,times}
\usepackage{natbib}
\usepackage{amssymb,amsmath}
\bibpunct{(}{)}{;}{a}{}{,}

\begin{document}

   \title{Structures of GMC W\,37}

   \volnopage{Vol.0 (200x) No.0, 000--000}      
   \setcounter{page}{1}          

   \author{Xiaoliang Zhan
      \inst{}
   \and Zhibo Jiang
      \inst{}
   \and Zhiwei Chen
      \inst{}
   \and Miaomiao Zhang
      \inst{}
   \and Chao Song
      \inst{}
   }
   \institute{Purple Mountain Observatory $\&$ Key Laboratory for Radio Astronomy, Chinese Academy of Sciences, 2 West Beijing Road, 210008 Nanjing, China; {\it zbjiang@pmo.ac.cn}\\
             }

   \date{Received September 15, 1996; accepted March 16, 1997}


  \abstract
   {We carried out observations toward the giant molecular cloud W\,37 with the $J = 1 - 0$ transitions of $^{12}$CO, $^{13}$CO, and C$^{18}$O using the 13.7\,m single-dish telescope at the Delingha station of Purple Mountain Observatory. Based on the three CO lines, we calculated the column densities, cloud masses for the molecular clouds with radial velocities at around $+20\,\mathrm{km\,s}^{-1}$. The gas mass of W\,37, calculated from $^{13}$CO emission, is $1.7\times10^5\,M_\odot$, above the criteria of giant molecular cloud. The dense ridge of W\,37 is a dense filament, which is supercritical in linear mass ratio. Dense clumps found by C$^{18}$O emission are aligned along the dense ridge with a regular interval about 2.8\,pc, similar to the clump separation caused by large-scale `sausage instability'. We confirm the identification of the giant molecular filament (GMF) G\,18.0-16.8 by \cite{2014A&A...568A..73R} and find a new giant filament, G\,16.5-15.8, located in the west $\sim0.8\degr$ of G\,18.0-16.8. Both GMFs are not gravitationally bound, as indicated by their low linear mass ratio ($\sim80\,M_\odot\,\mathrm{pc}^{-1}$). We compared the gas temperature map with the dust temperature map from \emph{Herschel} images, and find similar structures. The spatial distributions of class I objects and the dense clumps is reminiscent of triggered star formation occurring in the northwestern part of W\,37, which is close to NGC\,6611. 
}
   \authorrunning{X.L. Zhan et al. }            
   \titlerunning{Structures of the W 37 molecular cloud}  
   \maketitle
%

\section{Introduction}
\label{sect:intro}
 Giant molecular clouds (GMCs) are generally accompanied with star formation activities. In the close proximity of the cloud core region of GMC W\,37, the well-known star-forming region M\,16 or so-called `Eagle nebula' represents  the most active area of star formation inside W\,37. M\,16  is ionized by NGC\,6611, a massive young star cluster which is dominated by the massive binary system HD\,168076 consisting of an O3.5\,V and an O7.5\,V star   (\citealt{2005A&A...437..467E,2008A&A...489..459M,2008A&A...481L..99A}). This system has a total of $75-80$ solar masses and provides about half the ionizing radiation for the nebula (\citealt{2005A&A...437..467E}). The studies of NGC\,6611 show that there are numbers of OB members (\citealt{2008A&A...489..459M,2008A&A...481L..99A}), and the members are as young as only few million years (\citealt{1993AJ....106.1906H,2007A&A...462..245G}). The interplay between the OB cluster NGC\,6611 and the surrounding molecular clouds (part of W\,37) have been observed in various scales. For instance, the CO\,$1-0$ observations with the BIMA interferometer at a resolution about 0.1\,pc of the   `pillars of creation' (\citealt{2015MNRAS.450.1057M})  in M\,16 show that the velocity gradients along the  `pillars of creation' are produced by ionization front impact with a cloud core (\citealt{1998ApJ...493L.113P}). In the scale of the whole GMC W\,37, the \emph{Herschel} images enable \cite{2012A&A...542A.114H} to find a  prominent eastern filament running  southeast-northwest and away from the high-mass star-forming central region M\,16 and the NGC\,6611 cluster, as well as a northern filament which extends around and away from the cluster in the forms of dust temperature and column density maps. Moreover, the dust temperature in each of these filaments decreases with increasing distance from the NGC\,6611 cluster, indicating a heating penetration depth of $\sim10\,$pc in each direction in $3-6\times10^{22}\,\mathrm{cm}^{-2}$ column density filaments. These results suggest that the NGC\,6611 cluster impacts the temperature of future star-forming sites inside the W\,37 cloud, modifying the initial conditions for collapse (\citealt{2012A&A...542A.114H}). 

Previous CO lines observations toward the GMC W\,37 were from the $^{12}$CO\,$1-0$ survey of the entire Milky Way (\citealt{2001ApJ...547..792D}). This CO survey (hereafter 1.2\,m CO survey) has an angular resolution $\sim 8\farcm5$ at 115\,GHz, and a velocity resolution mostly at around 0.65\,km\,s$^{-1}$. Along the same line of sight (LOS) toward W\,37, the 1.2\,m CO survey detects CO line emissions in the range between $-5\,\mathrm{km\,s}^{-1}$ and $140\,\mathrm{km\,s}^{-1}$. \cite{2014A&A...568A..73R} identify a giant molecular filament (GMF), which is running  in the galactic longitude range $18.0\degr-16.8\degr$ with a length of 88\,pc and its near end located about only $15\arcmin$ away from the GMC W\,37. The velocity range of this GMF is $21-25\,\mathrm{km\,s}^{-1}$ in the form of $^{13}$CO\,$1-0$ line emissions, very close to the velocity of W\,37 ($\sim20\,\mathrm{km\,s}^{-1}$). Because of this coherent velocity structure, \cite{2014A&A...568A..73R} regard the association of this GMF with respect to W\,37.

In this work, we present the preliminary results of the CO\,$1-0$ line observations toward the galactic coordinate range $l=15.5-18.5\degr$ and $b=0.0-1.5\degr$, as a part of the new Galactic Plane survey using the $J= 1 - 0$ transitions of $^{12}$CO, $^{13}$CO, and C$^{18}$O molecules. In section 2, the observations and data reduction are described. The results based on the CO lines are presented in section 3. Section 4 discusses the structures of W 37. A summary is addressed in section 5.

\begin{table}
\bc
\caption[]{Observation Parameters\label{tab1}}
 \begin{tabular}[t]{c c c c c c c}
 \hline
 \hline
Line & v$_0$ & HPBW & T$_{sys}$ & $\eta_{mb}$ & $\delta_v$ & T$_{mb}$ rms noise \\
($J$ = 1 - 0) & (GHz) & ($''$) & (K) &   & (km s$^{-1}$)  & (K) \\
 \hline
$^{12}$CO & 115.271204 & 52$\pm$3  & 220-500 & 43.6\% & 0.160 & 0.55 \\
$^{13}$CO & 110.201353 & 52$\pm$3  & 150-310 & 48.0\% & 0.158 & 0.22 \\
C$^{18}$O & 109.782183 & 52$\pm$3  & 150-310 & 48.0\% & 0.158 & 0.22 \\
 \hline
 \end{tabular}
 \ec
 \tablecomments{0.92\textwidth}{The beam width and main beam effieiency are given by annual report of the telescope status in the year 2013.}
\end{table}

\section{Observations and data reduction}
\label{sect:Obs}
We observed  GMC W\,37 in  $^{12}$CO\,$1-0$, $^{13}$CO\,$1-0$, and C$^{18}$O\,$1-0$ with the Purple Mountain Observatory Delingha (PMODLH) 13.7\,m telescope (\citealt{2011ChA&A..35..439Z}) as one of the scientific targets regions of the Milky Way Imaging Scroll Painting (MWISP) project\footnote{http://www.radioast.nsdc.cn/yhhjindex.php}  (\citealt{2015ApJ...798L..27S}). The time span of the observations is from October 6th to December 13th, 2013. The three CO\,$1-0$ lines were observed simultaneously with the 9-beam Superconducting Spectroscopic Array Receiver  (\citealt{2012ITTST...2..593S}) working in sideband separation mode and with the Fast Fourier transform spectrometer . The pointing accuracy is checked every year by carrying out five-point observation for the planets (e.g., Jupiter), and found to be stably better than $5\arcsec$ when tracking targets. During all observations, standard sources were observed every two hours for calibration and monitoring the systematic performance.  The typical receiver noise temperature (T$_{rx}$) is about 30 K as given by the status report of PMODLH in the year 2013. The typical system temperature during observations is 280\,K for $^{12}$CO\,$1-0$ and 185\,K for $^{13}$CO\,$1-0$ and C$^{18}$O\,$1-0$.

\begin{figure}[!hbt]
\bc
\centering
   \includegraphics[width=4.9in]{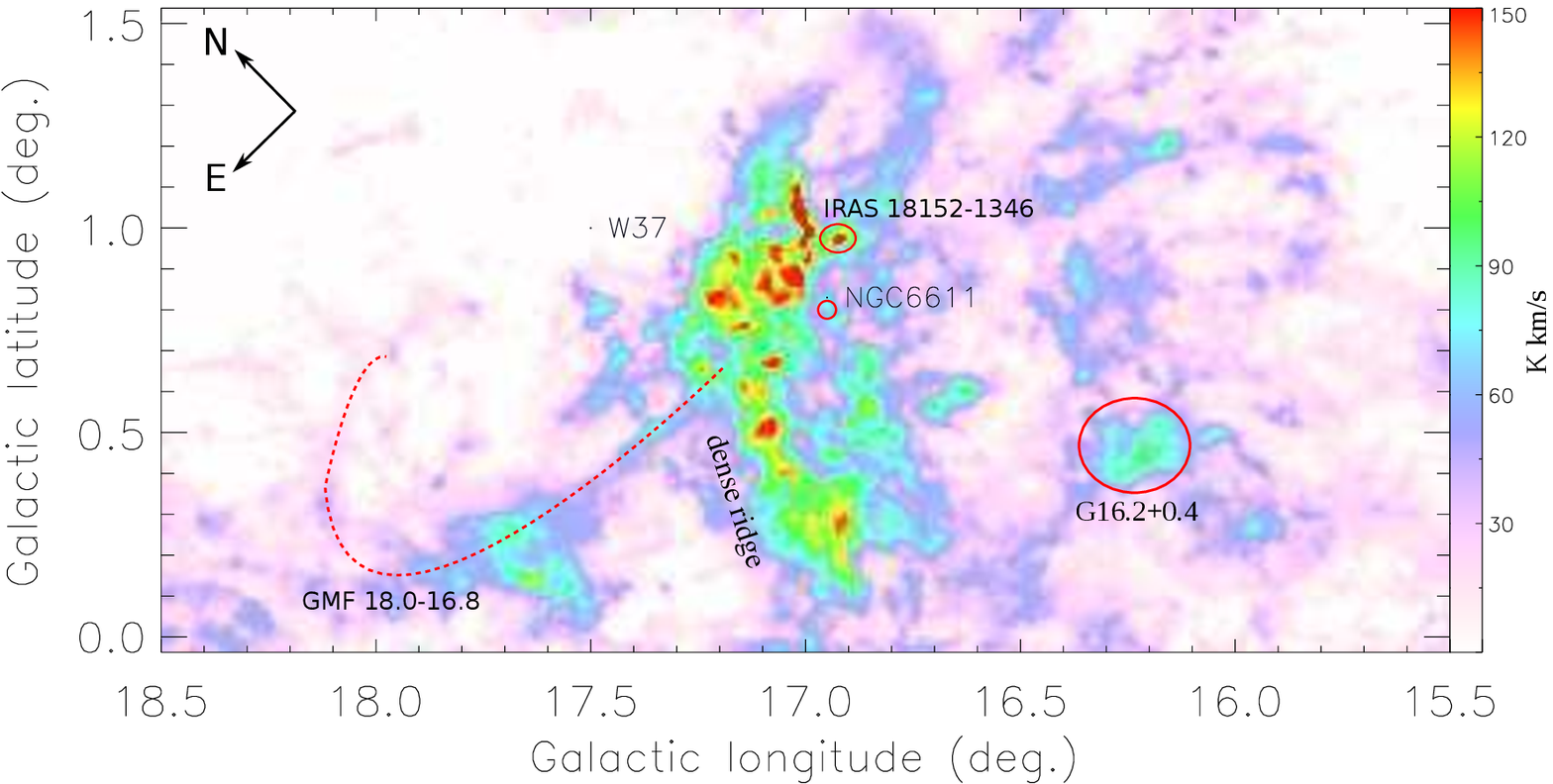}
   \includegraphics[width=4.9in]{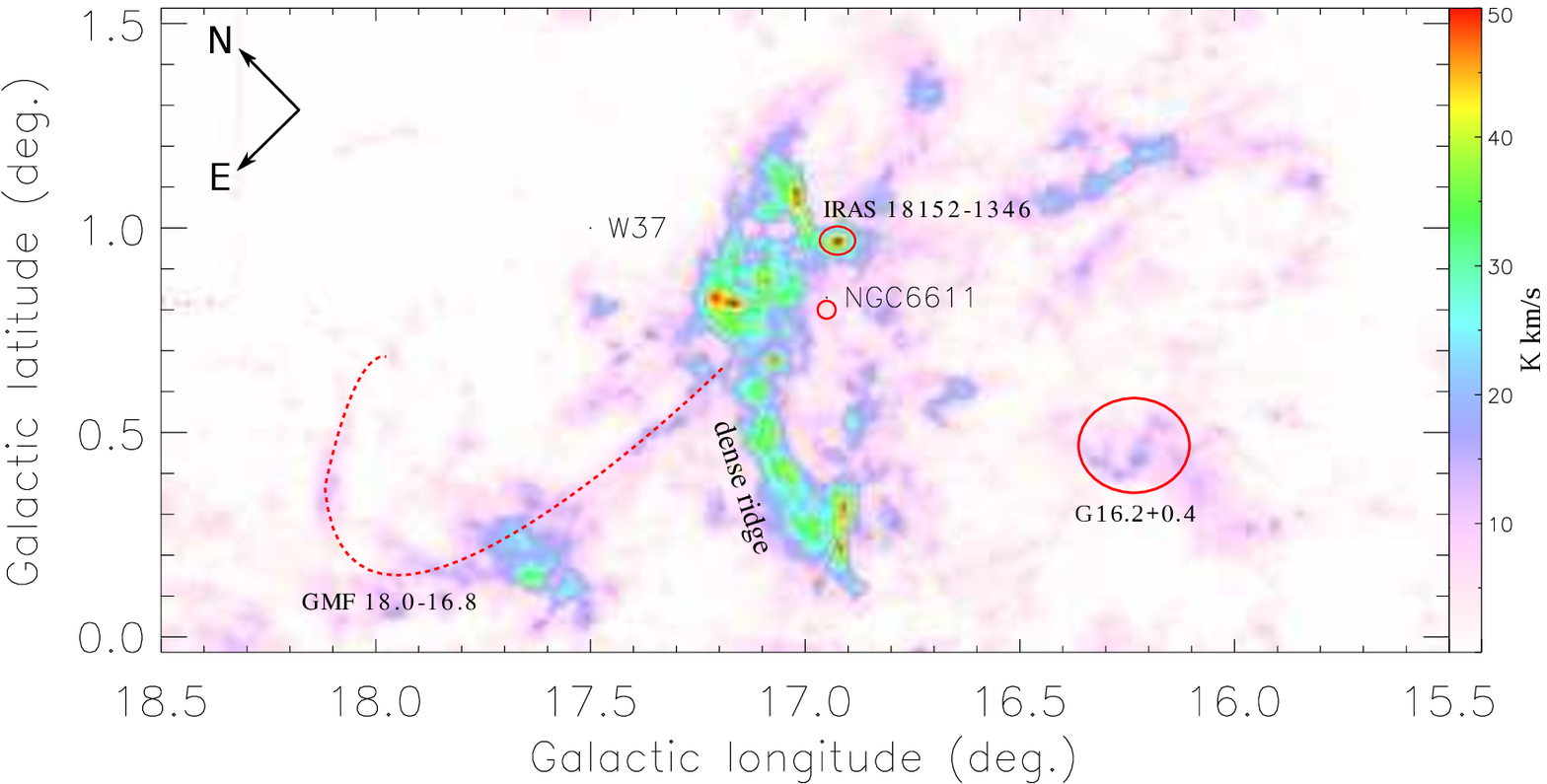}
   \includegraphics[width=4.9in]{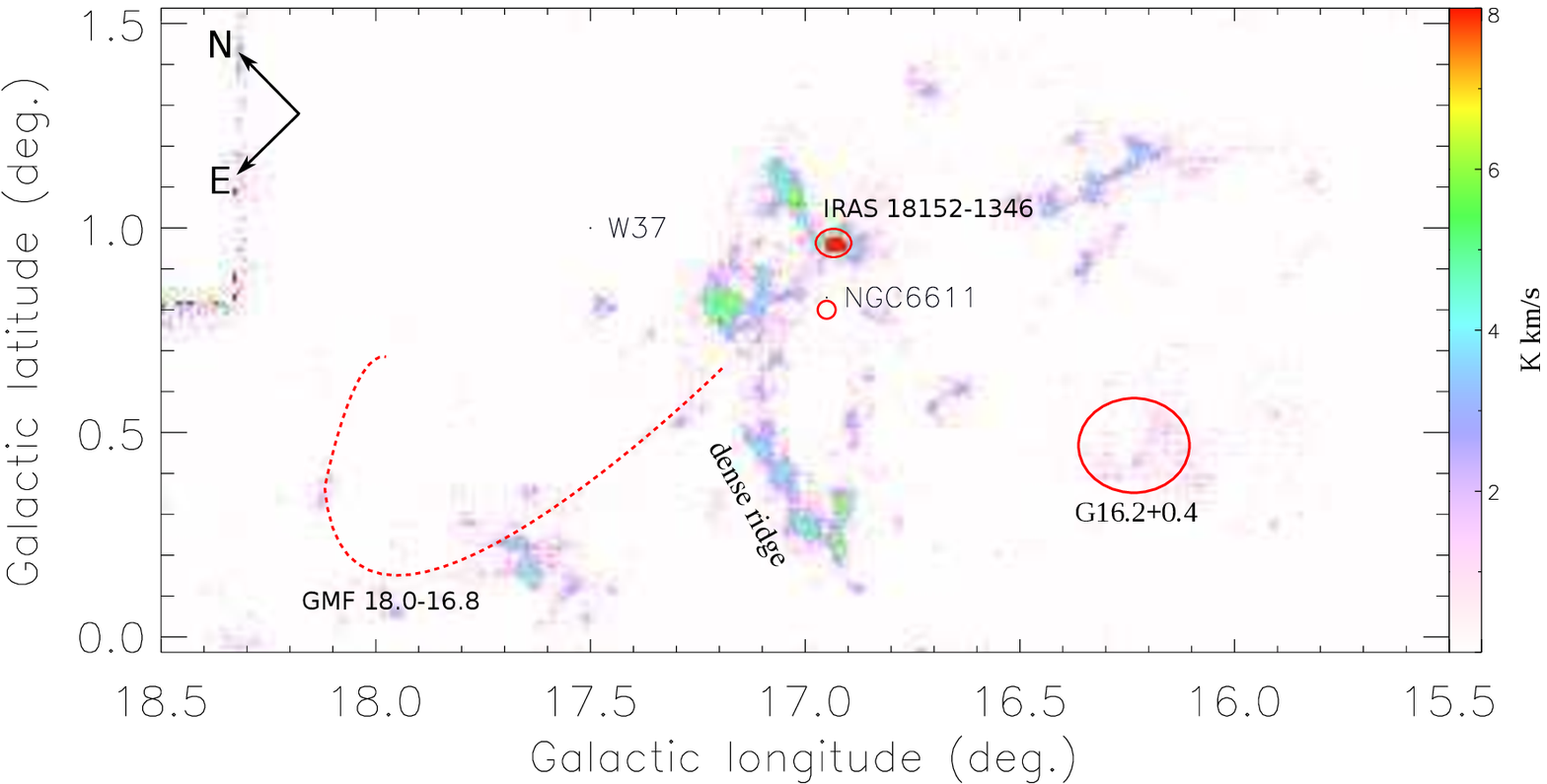}
 \caption{Top panel: the integrated intensity map of $^{12}$CO\,$1-0$ in the range $16-27\,\mathrm{km\,s}^{-1}$; middle: same as the top but for $^{13}$CO\,$1-0$; bottom: same as the top but for C$^{18}$O\,$1-0$. The large-scale filament G\,18.0-16.8 is outlined by the red dashed line. The locations of the high-mass young cluster NGC\,6611, IRAS 18152-1346, and isolated molecular cloud G16.2+0.4 are denoted by the red circles/ellipses.}
 \label{fig1}
\ec
\end{figure}

Our observations covered an area from galactic longitude $15.5\degr$ to $18.5\degr$ and galactic latitude from $0.0\degr$ to $1.5\degr$ (see Fig. \ref{fig1}). The GMC W\,37 was mapped using the on-the-fly observation mode, with the standard chopper wheel method for calibration (\citealt{1973ARA&A..11...51P}).  To conduct the observation, the whole region is split into many cells，each of $30\arcmin\times30\arcmin$ dimension. Each cell is scanned  along the galactic longitude and then the latitude direction on the sky at a constant rate of $50\arcsec$ per second, and receiver records spectra every 0.3 second. Every cell was repeatedly observed with this observation pattern until the root mean square (RMS) of spectra decreases down to the RMS levels required by MWISP, which are 0.5\,K for $^{12}$CO\,$1-0$ line, and 0.3\,K both for $^{13}$CO\,$1-0$ and C$^{18}$O\,$1-0$ line, respectively. 

After removing the bad channels in the spectra, we calibrated the antenna temperature ($Ta^*$) to the main beam temperature ($T_{mb}$) with a main beam efficiency of 44\% for $^{12}$CO and 48\% for $^{13}$CO and C$^{18}$O. The calibrated data were then re-gridded to 30$\arcsec$ pixels and mosaicked to a FITS cube using the GILDAS software package (\citealt{2000ASPC..217..299G}). A first order base-line was applied for the spectra. The resulting RMS level is 0.55 K for $^{12}$CO at the resolution of 0.159 km s$^{-1}$, 0.25 K for $^{13}$CO and 0.22 K for C$^{18}$O at the resolution of 0.166 km s$^{-1}$. A summary of the observation parameters is provided in Table~\ref{tab1}.

\begin{figure}[hbt]
\bc
\begin{minipage}[t]{0.5\linewidth}
\centering
\includegraphics[width=1\textwidth]{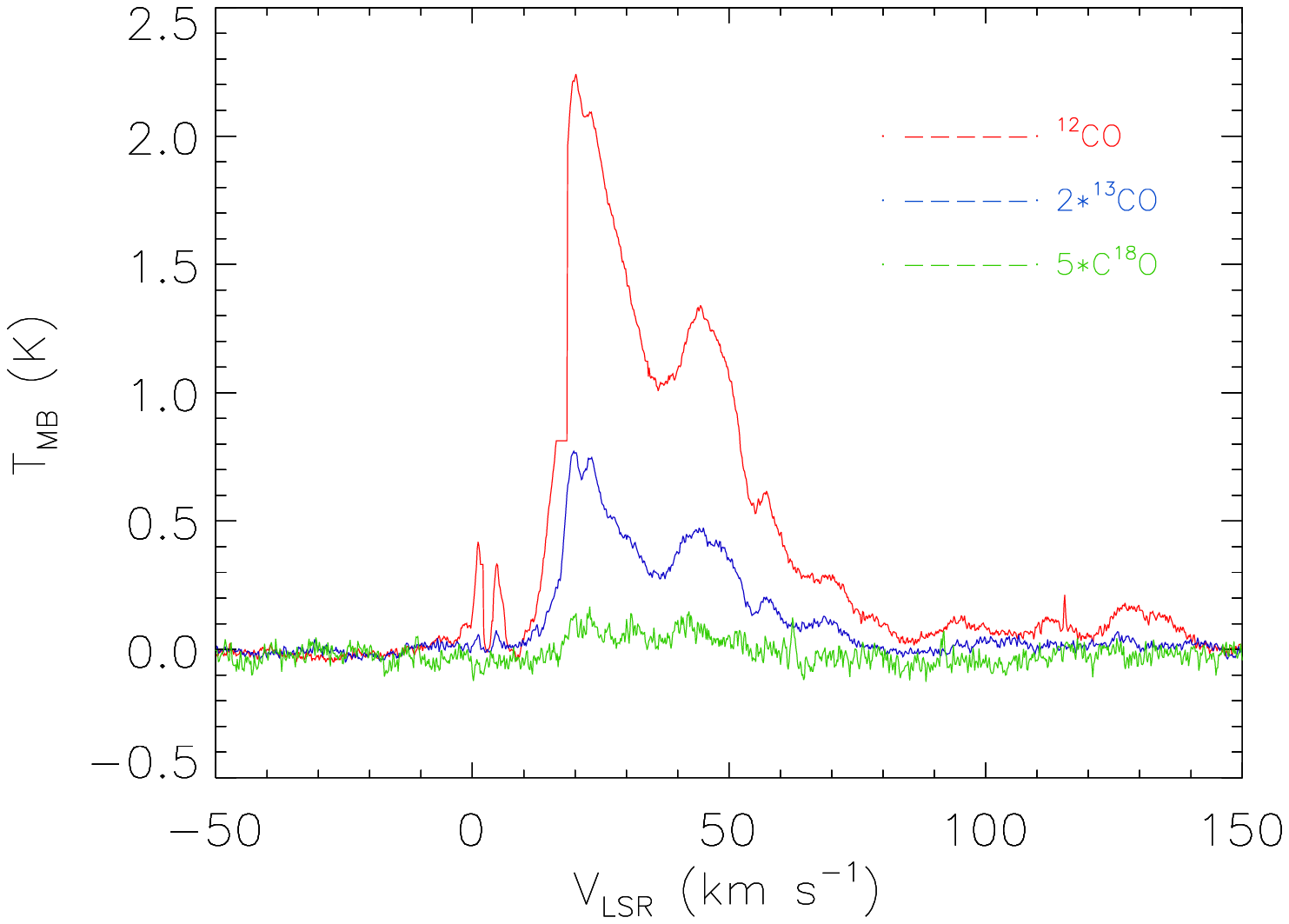}
\caption{Average spectra of the $^{12}$CO\,1$-$0, $^{13}$CO\,1$-$0, and C$^{18}$O\,1$-$0 lines of the observed region. }
\label{fig2}
\end{minipage}%
\begin{minipage}[t]{0.5\linewidth}
\centering
\includegraphics[width=1\textwidth]{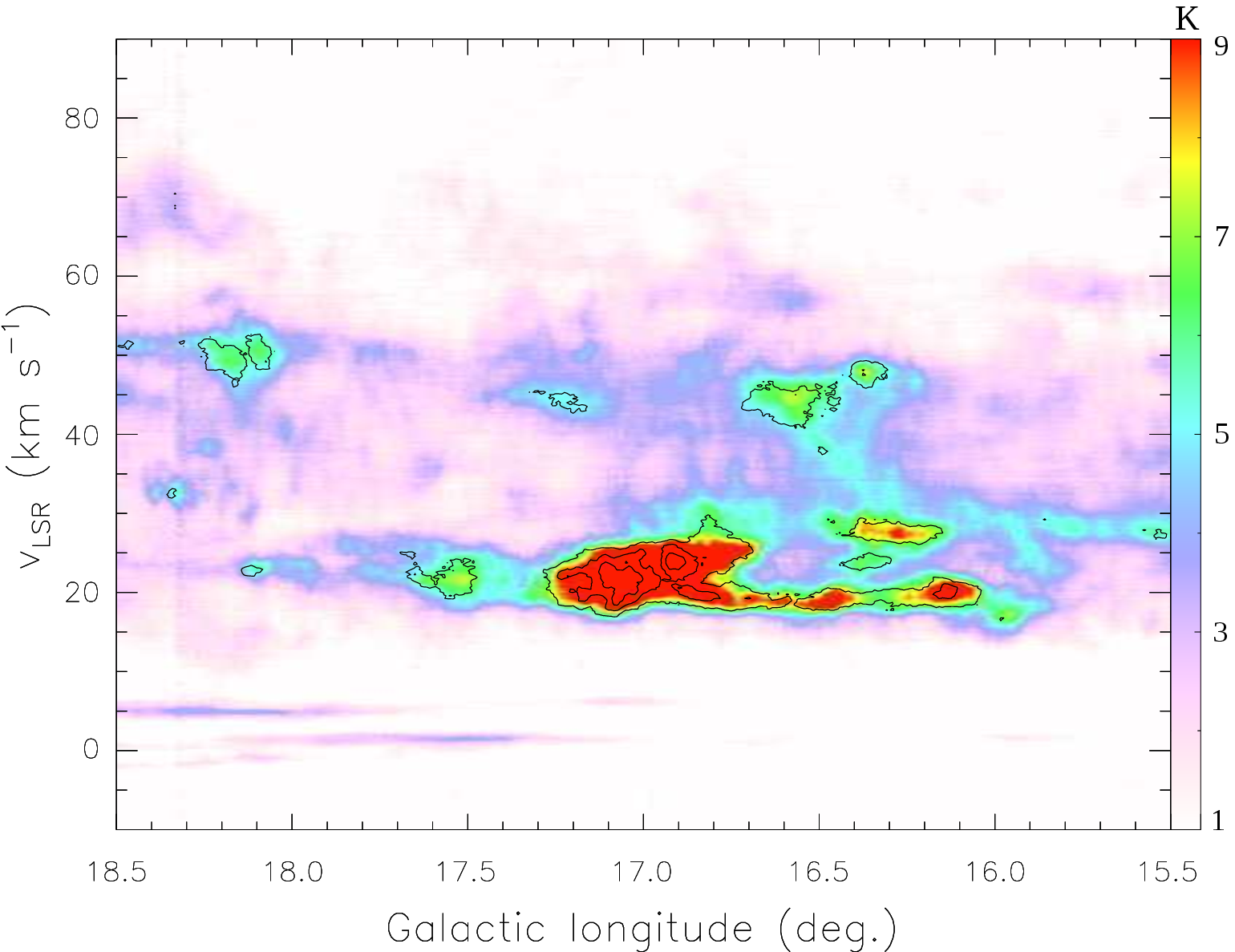}
\caption{Position-velocity map of $^{12}$CO\,1$-$0 emission in intensity (color scale) and $^{13}$CO\,1$-$0 emission in contours (black). The contours start from 1\,K with an interval of 1\,K. The CO emissions with radial velocities greater than $90\,\mathrm{km\,s}^{-1}$ are not shown because of their weaknesses. }
\label{fig3}
\end{minipage}
\ec
\end{figure}

%
\section{Results}
\label{sect:res}
\subsection{The basic characteristics of the CO emissions}
The average spectra of the whole observed region in the three lines are shown in Fig. \ref{fig2}. We clearly see several velocity components in the range $0-140\,\mathrm{km\,s}^{-1}$ in the $^{12}$CO\,$1-0$ spectrum; among them the component at around $20\,\mathrm{km\,s}^{-1}$ is strongest. Another component slightly weaker than the $20\,\mathrm{km\,s}^{-1}$ component shows a radial velocity at around $24\,\mathrm{km\,s}^{-1}$. Other velocity components are significantly weaker than the components of 20 km s$^{-1}$ and 24 km s$^{-1}$. Those diverse velocity components  likely come from different molecular clouds with different distances along the same  LOS.  Given the distances of the Galactic spiral arms from the Sun (\citealt{2008AJ....135.1301V}), as well as the rotation curve of the Galaxy (\citealt{2012PhRvD..85l4020M}), we can attribute  most of the observed CO emissions to their natal spiral arms. The components of W\,37 peaked at $20\,\mathrm{km\,s}^{-1}$ and $24\,\mathrm{km\,s}^{-1}$ have kinematic distances between $2.0-2.4$\,kpc. 
 Because of the association between M\,16 and W\,37, the distance of M\,16, 2.0\,kpc, used by \cite{2007ApJ...654..347L} is accepted here for W\,37's distance. W\,37 is located at the near side of the Sagittarius arm. We can marginally identify the molecular clouds located at the far side of the Sagittarius arm, which has typical radial velocity in the range $60-70\,\mathrm{km\,s}^{-1}$, as seen in Fig.~\ref{fig3}. The third strongest component at around $50\,\mathrm{km\,s}^{-1}$ is probably located at the Scutum-Centaurus Arm. Besides these molecular clouds on the spiral arms, there are other molecular clouds with radial velocities which cannot be attributed to any spiral arms lying on the same LOS. They might be inter-arm molecular clouds with relative low density.



\begin{figure}[!hbt]
\bc
  \includegraphics[width=1.0\textwidth]{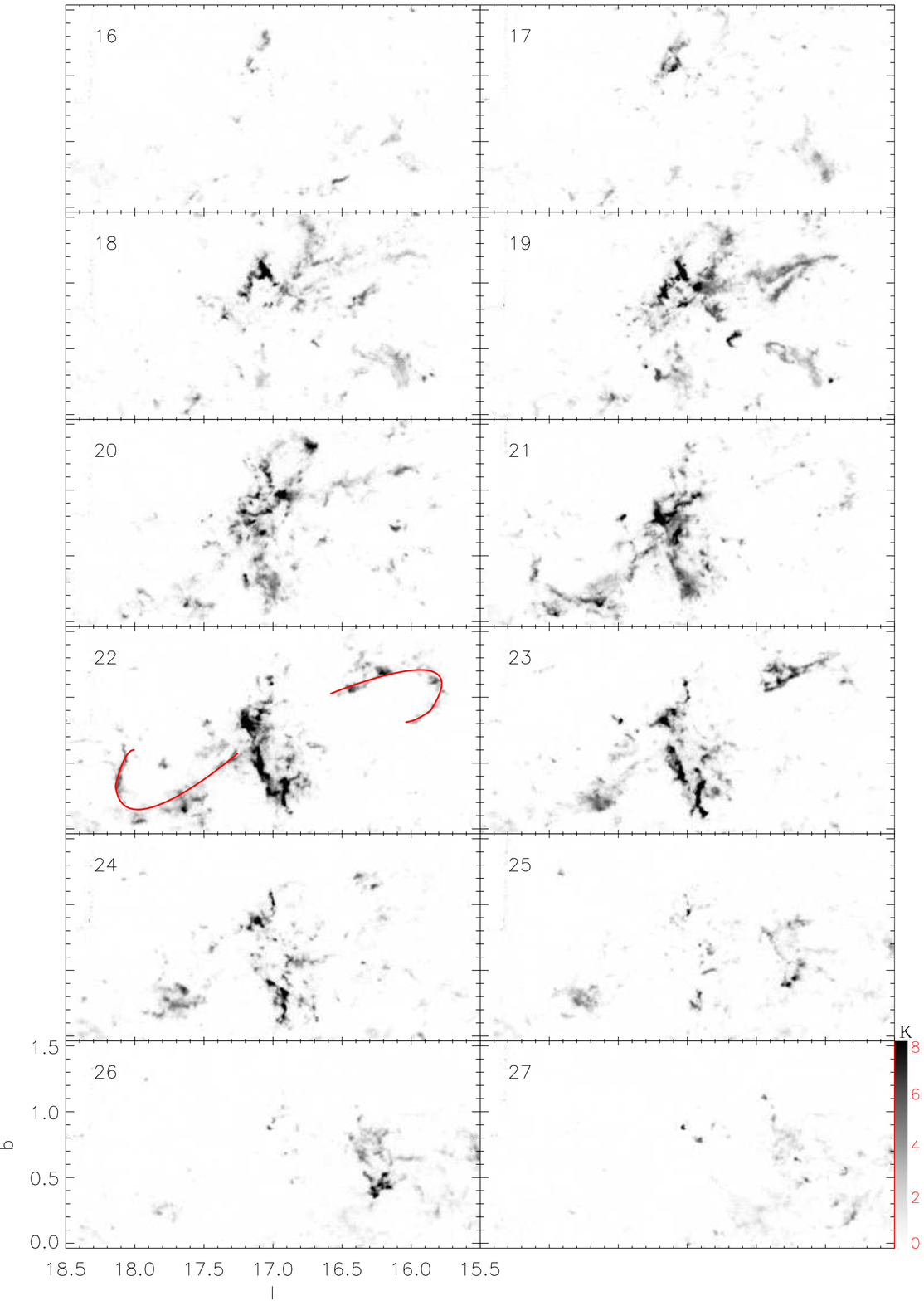}
  \caption{Channel map of $^{13}$CO\,$1-0$ emission of the observed region. Each frame is marked with the corresponding central velocity in km\,s$^{-1}$ on the upper-left corner. In the 21\,km\,s$^{-1}$ frame, the sketch curves in red denote the pair of filaments.}
  \label{fig4}
  \ec
\end{figure}

\begin{figure}[hbt]
\bc
\centering
\includegraphics[width=14.5cm, angle=0]{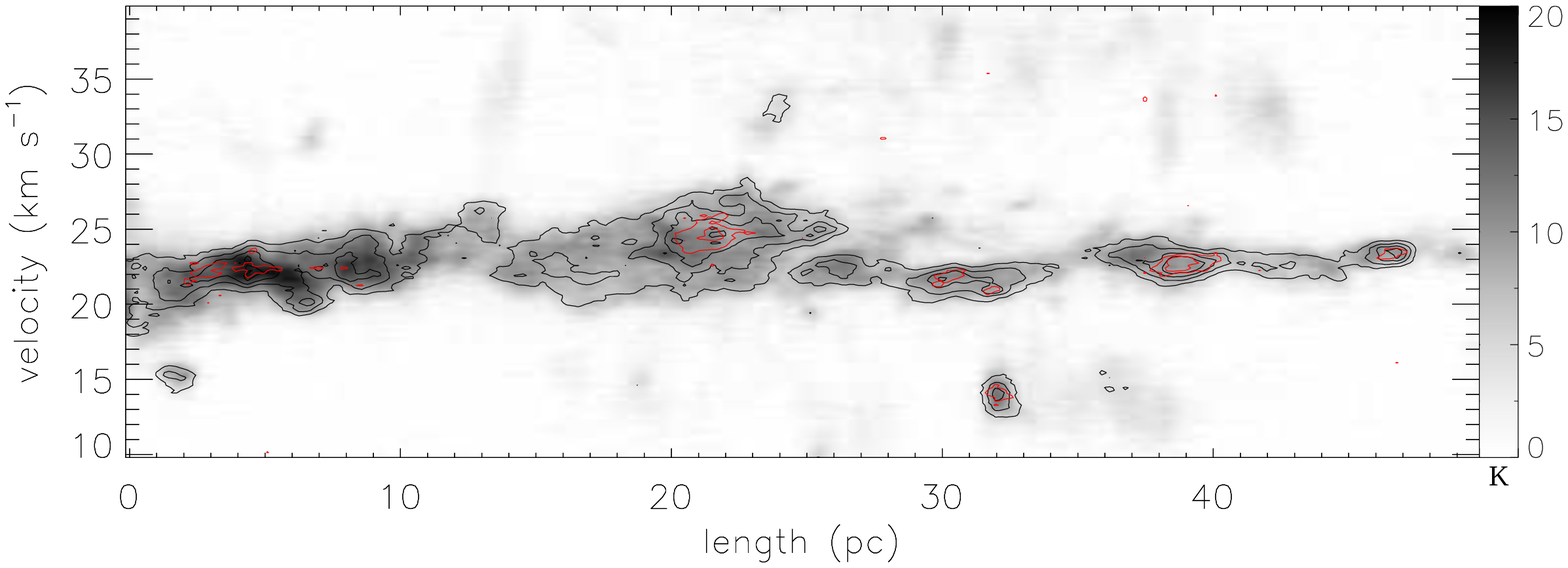}
\includegraphics[width=14.5cm, angle=0]{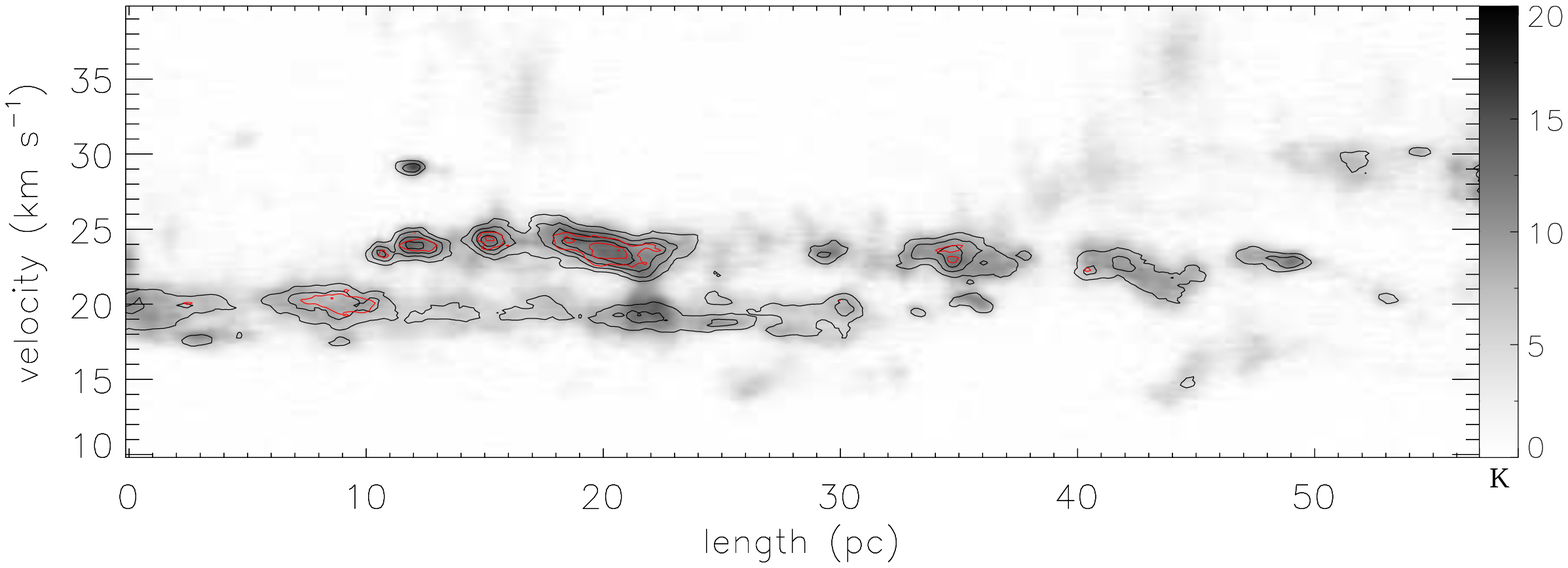}
\caption{Position-velocity maps of $^{12}$CO\,1$-$0 emission in intensity (grey scale) of the two GMFs, along the paths sketched by the red curves in Fig.~\ref{fig4}. The overlaid black contours are $^{13}$CO\,$1-0$ emission starting from 1.3\,K with an interval of 1.3\,K, and red contours are $^{18}$CO\,$1-0$ emission starting from 0.8\,K with an interval of 0.4\,K. The left part is shown in the top panel, and the right part in the bottom. In each panel, the length of each filament is counted from the end toward W\,37 side.}
\label{fig5}

\ec
\end{figure}

\subsection{The structure of W\,37}
This paper focuses on the bulk gas component of W\,37 with a peak radial velocity $\sim20\,\mathrm{km\,s}^{-1}$. Fig.~\ref{fig1} shows the integrated intensity maps of the three lines. The distribution of $^{12}$CO gas outlines the overall morphology of the GMC W\,37, which is running  perpendicular to the Galactic plane with a length about 46\,pc. While the $^{13}$CO\,$1-0$ emissions trace the relatively dense parts of W\,37, which also distribute perpendicular to the Galactic plane. In the C$^{18}$O\,$1-0$ emission, W\,37 can be split into two major parts: the dense filamentary structure in between $b=0.2-0.7\degr$ (hereafter the dense ridge) and the northwestern part closest to the massive cluster NGC\,6611. A very bright clump located in northwest of NGC\,6611 shows stronger C$^{18}$O emission than the dense ridge .  \cite{2007ApJ...666..321I} identified this bright clump as a massive young stellar object (YSO), also known as IRAS\,18152-1346.  

The column density map of W\,37 (37\arcsec resolution, Fig.~2 in \citealt{2012A&A...542A.114H}) derived from the \emph{Herschel} images shows very similar morphology to the $^{12}$CO gas distribution. However, what the dust column density map cannot tell us is the morphology variation of W\,37 along with the radial velocity in our CO lines observations. Therefore, we show this morphological variation in the form of $^{13}$CO\,$1-0$ channel map in Fig.~\ref{fig4}. The dense ridge appears mainly in the range $\sim20-23\,\mathrm{km\,s}^{-1}$, meanwhile the northwestern part shows wider velocity range from $17\,\mathrm{km\,s}^{-1}$ to $24\,\mathrm{km\,s}^{-1}$.    There are dense clumps  ($V_\mathrm{LSR}\gtrsim 25\,\mathrm{km\,s}^{-1}$) at positions around $l=16.2\degr, b=0.4\degr$, separated from the dense ridge about $0.7\degr$, or 25\,pc at a distance of 2.0\,kpc. We refer these dense clumps as the isolated molecular cloud G16.2+0.4, which is independent to the GMC W\,37. 

Interestingly, we note a pair of filaments located in the  two sides of W\,37 in the velocity range $20-23\,\mathrm{km\,s}^{-1}$. This pair of filaments is symmetric with respect to W\,37. The position velocity maps of this symmetric pair of filaments are shown in Fig.~\ref{fig5}.  This pair of filaments show coherent velocity distribution peaked both at around $23\,\mathrm{km\,s}^{-1}$. 

\subsection{Masses, column densities of the clouds}
Better than the 1.2\,m CO survey using a single CO line, our new CO three-line survey enables us to obtain the column density of CO gas under the assumption of local thermodynamic equilibrium (LTE). In general, the $^{12}$CO\,$1-0$ line is optically thick, hence the excitation temperature can be determined via the formula: 
\begin{equation}
  T_\mathrm{ex}=5.5/\ln(1+\frac{5.5}{T_\mathrm{B}(^{12}\mathrm{CO})+0.82}).
\label{eq:1}
\end{equation}
We assume that $^{13}$CO and C$^{18}$O have the same $T_\mathrm{ex}$ as $^{12}$CO. Assuming $^{13}$CO and C$^{18}$O lines optically thin, the optical depths of the two isotopes of $^{12}$CO are only functions of $T_\mathrm{B}$ and $T_\mathrm{ex}$:
\begin{equation}
  \tau_0=-\ln\left[1-\frac{T_B}{5.3}\left\{ [\exp(\frac{5.3}{T_{ex}}-1)]^{-1}-0.16\right\}^{-1}\right].
\label{eq:2}
\end{equation}
To obtain the column density for a molecule, i.e. $^{13}$CO or C$^{18}$O, one must sum over all energy levels of the molecule. In the LTE case, the total column density of $^{13}$CO or C$^{18}$O is given by 
\begin{equation}
  N(total)_\mathrm{CO}=3.0\times 10^{14} \frac{T_{ex} \int_{}^{} \tau (v) {\rm d}v}{1-\exp\{-5.3/T_{ex}\}}.
\label{eq:3}
\end{equation}
Substituting the $\tau(^{13}\mathrm{CO})$ and $\tau(\mathrm{C^{18}O})$ can calculate the column density of $^{13}$CO and C$^{18}$O, respectively. 
Adopting fractional abundances of H$_2/^{13}$CO$=7\times10^5$ and H$_2/$C$^{18}$O$= 7\times 10^{6}$ (\citealt{1995A&A...294..835C}), $N(^{13}$CO) and $N(\mathrm{C^{18}O})$ are then converted to the H$_2$ column density, respectively. Then we calculate the mass of the H$_2$ gas as
\begin{equation}
  M_{LTE}=\mu m_H N(H_2) S,
\label{eq:4}
\end{equation}
where $\mu=2.83$ is the average molecular weight, and S is the summed area of $^{13}$CO\,$1-0$ S/N ratios $>3$. Note that this criteria filters out many diffuse areas where only the $^{12}$CO lines are significant. Therefore the total masses of the molecular clouds calculated from the $^{13}$CO\,$1-0$ line can be treated as the lower limits. . Specially, we refer the molecular gas mass derived from the C$^{18}$O\,$1-0$ line as the dense gas mass because of the highest critical density of the C$^{18}$O molecule. The physical properties of the GMC W\,37 and the isolated cloud G16.2+0.4 are listed in Table~\ref{table:2}. W\,37's physical properties and length (46\,pc) all fall in the typical ranges for the GMCs previously identified (e.g., \citealt{2011ApJ...729..133M} and references therein). The comparison between W\,37 and G16.2+0.4 clearly states the physical differences between a GMC and a molecular cloud.

\begin{table}
\bc
\begin{minipage}[]{100mm}
\caption{Physical properties of the molecular clouds. \label{table:2}}
\end{minipage}
\begin{tabular}[t]{c c c c c c}
\hline
\hline
Cloud & Velocity & N(H$_2$) &  N(H$_2$) & Cloud  & Dense Gas\\
      & range    & $^{13}$CO\,$1-0$ & C$^{18}$O\,$1-0$ & mass  &  mass    \\
      &[km\,s$^{-1}$] & [ cm$^{-2}$] & [ cm$^{-2}$] &  [M$_{\odot}$] & [M$_{\odot}$]\\
\hline
W\,37 & $17-25$ &$1.1\times 10^{22}$ & $2.1\times 10^{22}$ & $1.7\times10^5$ & $1.0\times10^5$  \\
G16.2+0.4 &$25-27$ & $1.4\times 10^{21}$ &$4.4\times 10^{21}$ & $2.4\times10^3$ & $3.4\times10^2$ \\
\hline
\end{tabular}
\ec

\tablecomments{0.86\textwidth}{For both molecular clouds, mass calculation is made at a distance of 2.0\,kpc (\citealt{2007ApJ...654..347L}).  }

\end{table}

\section{Discussion}
\label{sect:discussion}

\subsection{Filamentary structures}\label{sect:GMF}
\begin{figure}
\bc
\centering
\includegraphics[width=3.5in,angle=90]{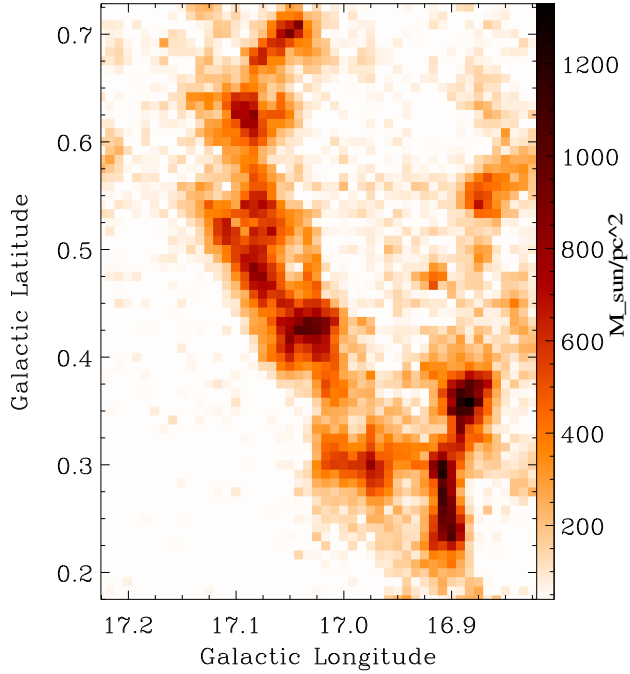}
\caption{Surface mass density map calculated from the $\mathrm{C^{18}O}$\,$1-0$ line emission for the dense ridge of W\,37.}
\label{fig6}
\ec
\end{figure}

The dense ridge shows a filament morphology between $b\approx0.25-0.72$. We show this peculiar part in Fig.~\ref{fig6} for clarifying its property. The dense ridge is clumpy, and clumps are distributed along the filament. The clumpy structure is most likely the result of gravitational collapse of a cylinder. If we regard the dense gas traced by the $\mathrm{C^{18}O}$\,$1-0$ emission as a whole, the stability of the filament can be described by the virial parameter $\alpha=M_\mathrm{vir}/M=2\sigma_v^2 l /GM$ (\citealt{2015ApJ...811..134S} and references therein), where $\sigma_v=\Delta_\mathrm{C18O}/2.355$, $l$, and $G$ are the average velocity dispersion of $\mathrm{C^{18}O}$\,$1-0$ emission, the length of the dense filament, and the gravitational constant, respectively. In the calculation, we take the long dense filament as a symmetrical cylinder. The geometric size of the cylinder ($\sim \mathrm{18\,pc}\times \mathrm{2\,pc} \times \mathrm{2\,pc}$) can be obtained from the $\mathrm{C^{18}O}$\,$1-0$ surface mass density map ($\sim 216 M_\odot~\mathrm{pc}^{-2}$ cutoff in Fig.~\ref{fig6}). We use the mean full width at half maximum (FWHM) of $\mathrm{C^{18}O}$\,$1-0$ emission ($\Delta_\mathrm{C18O}=1.9\,\mathrm{km\,s}^{-1}$) in the filament to estimate the mean velocity dispersion $\sigma_v=0.8\,\mathrm{km\,s}^{-1}$. The virial parameter $\alpha$ is estimated to be 0.6, indicating that the dense filament is gravitationally bound. Moreover, we can estimate the fragmentation separation within the filament to clarify whether `sausage instability' dominates the fragmentation process. Adopting $10^4\,\mathrm{cm}^{-3}$ as the mean density of the dense filament, the filament scale height $H=\sigma_v (4G\pi\rho_c)^{-0.5}$ is about 0.14\,pc due to the `sausage instability' of a self-gravitating fluid cylinder (\citealt{2015ApJ...811..134S} and references therein). This leads to a spacing of $22\,H\sim 3\,\mathrm{pc}$ between the fragmentation clumps. Applying the clump identification algorithm `GaussClumps' integrated in the Starlink/CUPID package~\footnote{More information about CUPID: http://starlink.eao.hawaii.edu/starlink/CUPID} to the surface mass density map of $\mathrm{C^{18}O}$\,$1-0$ emission, seven clumps are found along the dense filament. The seven clumps are roughly regularly spaced with a mean fragmentation separation about 2.8\,pc, which results in good agreement with the `sausage instability'. We notice another shorter and denser filament running in $b=0.2-0.4\degr$ in the right lower corner of Fig.~\ref{fig6}, which is worthy of detailed analyses in an individual paper.

We recall the pair of filaments almost perpendicular to W\,37, as noticed in the $^{13}$CO channel map Fig.~\ref{fig4}. The left part was identified by \cite{2014A&A...568A..73R} as a GMF named G\,18.0-16.8. However, our CO data reveal the counterpart of the G\,18.0-16.8, which is not found in \cite{2014A&A...568A..73R}'s paper. Upon the coherent velocity distribution of this pair of filaments, the GMF found by \cite{2014A&A...568A..73R} is only one part of this pair. Although the two parts of this pair are continuous in velocity, they are separated by a gap about 28\,pc, where W\,37 is located in between. The right half (hereafter G\,16.5-15.8) has a length about 50\,pc, at a distance of 2\,kpc. Moreover, the two GMFs have total masses on the order of $10^4\,M_\odot$, which are comparable to the masses of GMFs reported by \cite{2015MNRAS.450.4043W}. \cite{2014A&A...568A..73R} associate the left GMF with the GMC\,37 upon considering the comparable radial velocities between them. Because of the same argument, we take this association with respect to W\,37 for the right GMF. 

\begin{table}
\bc
\begin{minipage}[]{100mm}
\caption[]{Properties of GMFs\label{table:3}}
\end{minipage}
\begin{tabular}[t]{c c c c c c c c c}
\hline
\hline
Transition & \multicolumn{2}{c}{Mass ($\times$ 10$^{4}\,M_\odot$)} & \multicolumn{2}{c}{$N(\mathrm{H}_2)$ ($\times10^{21}\,\mathrm{cm}^{-2}$)} & \multicolumn{2}{c}{$\Delta v (\mathrm{km\,s}^{-1}$)}  \\
\cline{2-7}
Line & right & left & right & left & right & left \\
\hline 
 $^{13}$CO\,$1-0$ & 1.2 & 2.7 & 3.1 & 3.8 & 1.9 & 2.0  \\
 C$^{18}$O\,$1-0$ & 0.55 & 1.3 & 4.7& 5.6& 1.4 & 1.2  \\
\hline
\end{tabular}
\ec
\tablecomments{0.86\textwidth}{  For both molecular clouds, mass calculation is made at a distance of 2.0\,kpc (\citealt{2007ApJ...654..347L}). }
\end{table}

The geometric structure that the two GMFs are separated by W\,37 might be the result of the dynamical evolution of an entire GMF, which is disconnected by the contraction of W\,37 located at its middle way. In addition, about six class I objects have positions overlapped with the GMF on the left side, while no class I object is lying on the GMF on the right side (see Fig.~\ref{fig8}). However, this differential distribution between the two GMFs is not noticed for the class II objects, which seem to be equally distributed on the two GMFs. The differential distributions of class I objects on the two GMFs could be simply due to their differential physical properties. We can estimate the linear mass density (mass per unit length) of the two GMFs, which are help to clarify whether they are gravitationally stable or not. The mean linear mass density is thus estimated to be $88\,M_\odot~\mathrm{pc}^{-1}$ and $73\,M_\odot~\mathrm{pc}^{-1}$ for the left and right GMF, respectively, where gas mass is calculated from the $\mathrm{^{13}CO}$\,$1-0$ emission, filament length is obtained from Fig.~\ref{fig5}, and mean filament width is derived by dividing the filament area with filament length. The linear mass density has a maximum value $M_\mathrm{vir}/l=465 (\frac{\sigma}{\mathrm{km\,s^{-1}}})^2 M_\odot~\mathrm{pc}^{-1}$, over which a filament will break up into pieces.The mean FWHM of $\mathrm{^{13}CO}$\,$1-0$ emission (see Table~\ref{table:3}) leads to a maximum linear mass density about $300  M_\odot~\mathrm{pc}^{-1}$ for the two GMFs. Both GMFs are not dense enough to initiate large-scale fragmentation.   On the other hand, the mean dense gas fraction $M_\mathrm{C18O}/M_\mathrm{13CO}$ of the two GMFs are similar, which are 0.48 and 0.46 for the left and right GMF, respectively. To conclude, the mean physical properties of the two GMFs are very similar (see also Table~\ref{table:3}), which is against the differential distributions of class I objects. Alternatively, a more plausible explanation is class I objects are just overlapped with the left GMF along the same LOS with no physical associations.

\begin{figure}[ht]
\bc
  \centering
  \includegraphics[width=4.5in]{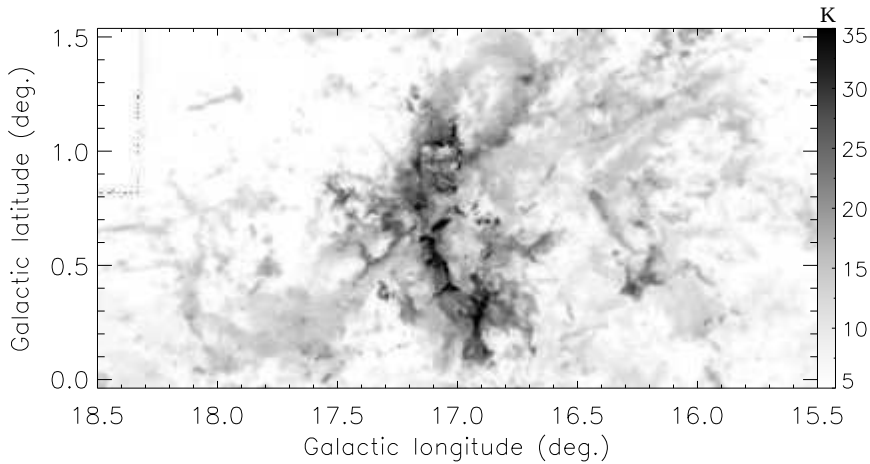}
  \caption{Gas excitation temperature map of the observed region, derived from the radiation temperature of $^{12}$CO\,$1-0$ emission in the range $16-27\,\mathrm{km\,s}^{-1}$.}
  \label{fig7}

\ec
\end{figure}

\begin{figure}[hbt]
\bc
  \includegraphics[width=4.5in]{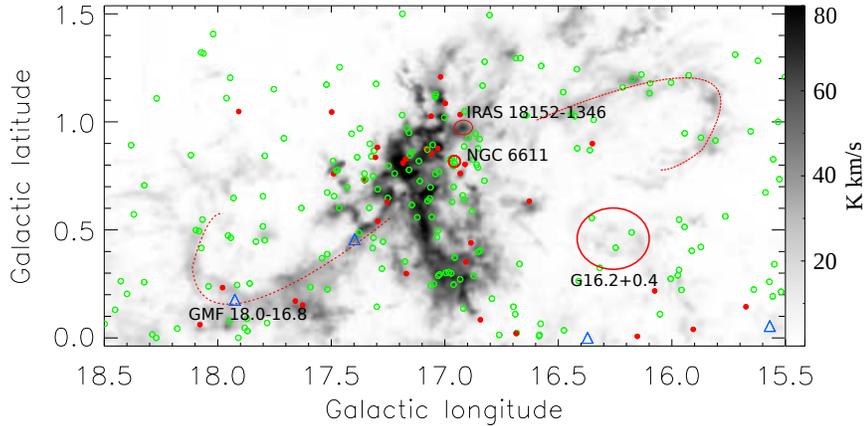}
  \caption{The distribution of YSO candidates projected inside the observed region. The red circles represent class I objects, green circles for class II objects, and blue triangles for transition disk objects. The two GMFs are outlined by the red dash lines. The locations of the high-mass young cluster NGC\,6611, IRAS 18152-1346, and isolated molecular cloud G16.2+0.4 are denoted by the red circles/ellipses. The background is the integrated intensity map of $^{12}$CO\,$1-0$.}
  \label{fig8}
\ec  
\end{figure}


\subsection{Young stellar objects in W\,37}
Fig.~\ref{fig7} shows the excitation temperature map of GMC W\,37. There are two domains of $T_\mathrm{ex}\gtrsim30$, which correspond to the dense ridge and the northwestern part of W\,37, respectively. These relative warm clouds are believed to be at the very beginning phases of gravitational contraction.  Moreover, the dense ridge and the northwestern part of W\,37 are also prominent in the dust temperature map from \emph{Herschel} images (\citealt{2012A&A...542A.114H}), whereas, they both shows dust temperature about 17\,K, lower than the gas temperature. This temperature difference might naturally comes from the two temperature tracers. The detected $^{12}$CO\,$1-0$ emission mostly originates from the surface layer because the emission from inner areas are self-absorbed, however, the dust continuum emission even from the inner most area can also be detected. In addition, this temperature difference is consistent with the theoretical point that the surface layer is warmer than the inner area of a molecular cloud.

 We used the latest WISE (\citealt{2010AJ....140.1868W}) source catalogue ALLWISE~\footnote{ALLWISE catalogue can be retrieved by http://wise2.ipac.caltech.edu/docs/release/allwise/} to identify YSO candidates located inside the observed region, adopting the criteria from \cite{2014JMP....55l2202K}. The class I objects are mostly distributed in the proximity of molecular gas, while the class II objects can be also frequently seen in areas with little/no molecular gas (see Fig.~\ref{fig8}). This phenomenon can be interpreted as the result of the age differences between class I and class II objects. Further on, class I objects are better tracers of early stage star formation than class II objects. We still note that the class II objects have a trend of being concentrated toward the GMC W\,37, and much more class I objects are found in the northwestern part than in the dense ridge.  The class I objects in the northwestern part of W\,37 are located close to NGC\,6618. \cite{2012A&A...542A.114H} argued that NGC\,6611 impacts the conditions of W\,37 by heating. The penetrating depth of this heating effect is found to be about 9\,pc for the northwestern part whose column density is $2.3\times10^{22}\,\mathrm{cm^{-2}}$ (\citealt{2012A&A...542A.114H}). This penetration depth is almost identical to the northwestern part's size, indicating  that molecular clumps inside the northwestern part are very likely influenced by the cluster's heating effect. More class I objects found in the area close to NGC\,6611 can be interpreted as the result of the heating effect of the cluster, i.e., NGC\,6611 triggers the star formation in the northwestern part of W\,37. Because the northern part of W\,37 is massive and dense enough to avoid of being destroyed by the high-mass cluster NGC\,6611, the star formation inside this area can be the result of the direct compression of pre-existing density enhancements in this area (\citealt{1998ASPC..148..150E}). Therefore, the northwestern part of W\,37 is a promising candidate of triggering star formation by cluster's influence, and is worthy of continuous investigation.

\section{Summary}
\label{sect:summary}
    We observed the $J = 1 - 0$ transition lines of $^{12}$CO, $^{13}$CO, and C$^{18}$O toward GMC W\,37, as a part of the MWISP project. The bulk of W\,37's molecular gas have radial velocities between $17-24\,\mathrm{km\,s}^{-1}$, which can be attributed to the near end of the Sagittarius arm. The gas mass traced by $^{13}$CO emission is $1.7\times10^5\,M_{\odot}$ for W\,37. The dense ridge of W\,37 is a dense and gravitationally bound filament, as indicated by its large linear mass ratio. Dense clumps traced by C$^{18}$O emission are regularly spaced along the dense ridge with a mean separation about 0.28\,pc, which agrees with the mean clump separation caused by `sausage instability' in large scale.  Our comprehensive CO lines survey toward W\,37 confirm the discovery of G\,18.0-16.8 by \cite{2014A&A...568A..73R}, and identify another GMF, G\,16.5-15.8, in the west $0.7\degr$ of G\,18.0-16.8. Both GMFs have very similar linear mass ratio about $80\,M_\odot\,\mathrm{pc}^{-1}$, far less than the maximum value $300\,M_\odot~\mathrm{pc}^{-1}$, over which a filament will break up into pieces.  The gas excitation temperature map shows that W\,37's dense ridge and northwestern part have higher temperatures than the ambient gas, in contrast with the dust temperature derived by \cite{2012A&A...542A.114H}.  Combining the spatial distributions of class I objects identified by AllWISE catalog and the dense clumps from our CO emission maps, it is likely that the star formation activities within the W\,37 northwestern part are triggered by the associated nearby massive young cluster NGC\,6611. 

\begin{acknowledgements}
This work is supported by the Strategic Priority Research Program `The Emergence of Cosmological Structure' of the Chinese Academy of Sciences, grant No. XDB09000000, the Millimeter Wave Radio Astronomy Database, and the Key Laboratory for Radio Astronomy, CAS. Z.J. acknowledges the support by NSFC 11233007. X.Z. acknowledges Dr S. Zhang and Dr. Y. Su for their fruitful comments. This publication makes use of data products from the Wide-field Infrared Survey Explorer, which is a joint project of the University of California, Los Angeles, and the Jet Propulsion Laboratory/California Institute of Technology, funded by the National Aeronautics and Space Administration.
\end{acknowledgements}

\bibliographystyle{raa}
\bibliography{myrefs}

\begin{thebibliography}{26}
\providecommand{\natexlab}[1]{#1}
\providecommand{\selectlanguage}[1]{\relax}

\bibitem[{{Alecian} et~al.(2008){Alecian}, {Wade}, {Catala}
  et~al.}]{2008A&A...481L..99A}
{Alecian}, E., {Wade}, G.~A., {Catala}, C., et~al. 2008, \aap, 481, L99

\bibitem[{{Castets} \& {Langer}(1995)}]{1995A&A...294..835C}
{Castets}, A., \& {Langer}, W.~D. 1995, \aap, 294, 835

\bibitem[{{Dame} et~al.(2001){Dame}, {Hartmann}, \&
  {Thaddeus}}]{2001ApJ...547..792D}
{Dame}, T.~M., {Hartmann}, D., \& {Thaddeus}, P. 2001, \apj, 547, 792

\bibitem[{{Elmegreen}(1998)}]{1998ASPC..148..150E}
{Elmegreen}, B.~G. 1998, in Origins, \emph{Astronomical Society of the Pacific
  Conference Series}, vol. 148, edited by C.~E. {Woodward}, J.~M. {Shull}, \&
  H.~A. {Thronson}, Jr., 150

\bibitem[{{Evans} et~al.(2005){Evans}, {Smartt}, {Lee}
  et~al.}]{2005A&A...437..467E}
{Evans}, C.~J., {Smartt}, S.~J., {Lee}, J.-K., et~al. 2005, \aap, 437, 467

\bibitem[{{Guarcello} et~al.(2007){Guarcello}, {Prisinzano}, {Micela}
  et~al.}]{2007A&A...462..245G}
{Guarcello}, M.~G., {Prisinzano}, L., {Micela}, G., et~al. 2007, \aap, 462, 245

\bibitem[{{Guilloteau} \& {Lucas}(2000)}]{2000ASPC..217..299G}
{Guilloteau}, S., \& {Lucas}, R. 2000, in Imaging at Radio through
  Submillimeter Wavelengths, \emph{Astronomical Society of the Pacific
  Conference Series}, vol. 217, edited by J.~G. {Mangum} \& S.~J.~E. {Radford},
  299

\bibitem[{{Hill} et~al.(2012){Hill}, {Motte}, {Didelon}
  et~al.}]{2012A&A...542A.114H}
{Hill}, T., {Motte}, F., {Didelon}, P., et~al. 2012, \aap, 542, A114

\bibitem[{{Hillenbrand} et~al.(1993){Hillenbrand}, {Massey}, {Strom}, \&
  {Merrill}}]{1993AJ....106.1906H}
{Hillenbrand}, L.~A., {Massey}, P., {Strom}, S.~E., \& {Merrill}, K.~M. 1993,
  \aj, 106, 1906

\bibitem[{{Indebetouw} et~al.(2007){Indebetouw}, {Robitaille}, {Whitney}
  et~al.}]{2007ApJ...666..321I}
{Indebetouw}, R., {Robitaille}, T.~P., {Whitney}, B.~A., et~al. 2007, \apj,
  666, 321

\bibitem[{{Koenig} \& {Smolin}(2014)}]{2014JMP....55l2202K}
{Koenig}, R., \& {Smolin}, J.~A. 2014, Journal of Mathematical Physics, 55,
  122202

\bibitem[{{Linsky} et~al.(2007){Linsky}, {Gagn{\'e}}, {Mytyk}, {McCaughrean},
  \& {Andersen}}]{2007ApJ...654..347L}
{Linsky}, J.~L., {Gagn{\'e}}, M., {Mytyk}, A., {McCaughrean}, M., \&
  {Andersen}, M. 2007, \apj, 654, 347

\bibitem[{{Mannheim} \& {O'Brien}(2012)}]{2012PhRvD..85l4020M}
{Mannheim}, P.~D., \& {O'Brien}, J.~G. 2012, \prd, 85, 124020

\bibitem[{{Martayan} et~al.(2008){Martayan}, {Floquet}, {Hubert}
  et~al.}]{2008A&A...489..459M}
{Martayan}, C., {Floquet}, M., {Hubert}, A.~M., et~al. 2008, \aap, 489, 459

\bibitem[{{McLeod} et~al.(2015){McLeod}, {Dale}, {Ginsburg}
  et~al.}]{2015MNRAS.450.1057M}
{McLeod}, A.~F., {Dale}, J.~E., {Ginsburg}, A., et~al. 2015, \mnras, 450, 1057

\bibitem[{{Murray}(2011)}]{2011ApJ...729..133M}
{Murray}, N. 2011, \apj, 729, 133

\bibitem[{{Penzias} \& {Burrus}(1973)}]{1973ARA&A..11...51P}
{Penzias}, A.~A., \& {Burrus}, C.~A. 1973, \araa, 11, 51

\bibitem[{{Pound}(1998)}]{1998ApJ...493L.113P}
{Pound}, M.~W. 1998, \apjl, 493, L113

\bibitem[{{Ragan} et~al.(2014){Ragan}, {Henning}, {Tackenberg}
  et~al.}]{2014A&A...568A..73R}
{Ragan}, S.~E., {Henning}, T., {Tackenberg}, J., et~al. 2014, \aap, 568, A73

\bibitem[{{Shan} et~al.(2012){Shan}, {Yang}, {Shi}
  et~al.}]{2012ITTST...2..593S}
{Shan}, W., {Yang}, J., {Shi}, S., et~al. 2012, IEEE Transactions on Terahertz
  Science and Technology, 2, 593

\bibitem[{{Su} et~al.(2015){Su}, {Zhang}, {Shao}, \&
  {Yang}}]{2015ApJ...811..134S}
{Su}, Y., {Zhang}, S., {Shao}, X., \& {Yang}, J. 2015, \apj, 811, 134

\bibitem[{{Sun} et~al.(2015){Sun}, {Xu}, {Yang} et~al.}]{2015ApJ...798L..27S}
{Sun}, Y., {Xu}, Y., {Yang}, J., et~al. 2015, \apjl, 798, L27

\bibitem[{{Vall{\'e}e}(2008)}]{2008AJ....135.1301V}
{Vall{\'e}e}, J.~P. 2008, \aj, 135, 1301

\bibitem[{{Wang} et~al.(2015){Wang}, {Testi}, {Ginsburg}
  et~al.}]{2015MNRAS.450.4043W}
{Wang}, K., {Testi}, L., {Ginsburg}, A., et~al. 2015, \mnras, 450, 4043

\bibitem[{{Wright} et~al.(2010){Wright}, {Eisenhardt}, {Mainzer}
  et~al.}]{2010AJ....140.1868W}
{Wright}, E.~L., {Eisenhardt}, P.~R.~M., {Mainzer}, A.~K., et~al. 2010, \aj,
  140, 1868

\bibitem[{{Zuo} et~al.(2011){Zuo}, {Li}, {Sun} et~al.}]{2011ChA&A..35..439Z}
{Zuo}, Y.-X., {Li}, Y., {Sun}, J.-X., et~al. 2011, \caa, 35, 439

\end{thebibliography}
\end{document}